\documentclass[10pt]{article}
\usepackage{amsmath, a4wide, amssymb, mathrsfs, mcite, bbm}
\usepackage[english]{babel}
\usepackage[mathcal]{eucal}

\newcommand{\Hilbert}{\ensuremath{\mathscr{H}}}
\newcommand{\PHilbert}{\ensuremath{\mbox{P}\mathscr{H}}}

\title{{\bfseries On Arguments for Linear Quantum Dynamics}}
\author{M. Holman \\ Department of Physics and Astronomy, \hspace{0.05cm} Utrecht University, \\ Princetonplein 5,  \hspace{0.05cm} 3584 CC Utrecht, \hspace{0.05cm} The Netherlands \\ e-mail : {\ttfamily m.holman@phys.uu.nl}}

\begin{document}
\maketitle 
\begin{abstract}
\noindent Two recent arguments for linear dynamics in quantum theory are critically
re-examined. Neither argument is found to be satisfactory as it stands, 
although an improved version of one of the arguments can in fact be given. This improved
version turns out to be still not completely unproblematic, but it is argued that 
it contains only a single actual loophole, which is identical to a loophole that remains
in experimental proofs of nonlocality of Bell-type.
It is concluded that - within the context of the standard quantum kinematical framework 
and in agreement with what has been concluded by earlier authors - a nonlinear 
dynamics of density operators is inconsistent with relativistic causality.
However, it is also stressed that this conclusion in itself 
has little implication for the nature of dynamics at the Hilbert space level - 
in particular, it does not force dynamics to be linear at this level - nor does 
it continue to be valid in contexts that go beyond the standard quantum kinematical 
framework. Despite their seeming triviality, these last two points have not
always been appreciated in the literature. Finally, it is also pointed out that the 
argument for complete positivity, as given in conjunction with one of the two 
recent arguments for linear dynamics, in fact only establishes a condition 
that is weaker than complete positivity. 
\end{abstract}

\section{Introduction}
Given the standard kinematical framework for quantum theory (or the ``minimal'' 
extension of this framework\footnote{The standard way of representing states 
and observables in quantum theory, together with the squared modulus rule for 
calculating probabilities - or, equivalently, the trace rule - is here referred to 
as the (standard) kinematical framework for quantum theory. It is sometimes convenient to slightly generalize this standard 
framework, by allowing states to be represented by general density operators, while keeping 
the rest of the framework fixed. The resulting structure will occasionally be referred to as the 
minimally extended kinematical framework.}), it is natural to ask whether it's 
possible to consistently generalize the standard unitary dynamics of the theory, 
or whether perhaps this dynamics is already fixed by the kinematics, together 
with certain other, physically reasonable assumptions.\\ 
To this end it is recalled that in ordinary quantum theory, as well as in classical mechanics, dynamical evolution
is usually viewed as being generated by a preferred observable, i.e. the Hamiltonian.
Thus, within the context of the standard quantum kinematical framework (or its minimal
extension), in which the Hamiltonian is represented by a self-adjoint linear
operator, it seems at first that dynamics can only be unitary. Yet, an
important reason for considering non-standard dynamical evolution laws for
quantum theory comes from the existence of the projection postulate, which is 
assumed in most textbook formulations of the theory and which appears to introduce 
a second kind of dynamics into it. One is thus forced to either reject
this postulate, in which case it becomes necessary to work out a specific,
convincing interpretation of quantum theory, in which this postulate is not
needed (e.g. a nonlocal hidden variable interpretation), or to accept it, in 
which case overall conceptual consistency would
seem to require a demonstration that both types of dynamics can be viewed as 
different limits of a single dynamics, that arises within the context of a deeper
underlying theory\footnote{Alternatively, projections could be attributed to
``free choices'' made by the Universe, that are not associated with an underlying 
equation of motion - e.g. a stochastic modification of the Schr\"odinger equation.
A third option is discussed in section \ref{linearityargument1}, but in this case
severe interpretational difficulties arise.}.
If the second route is followed, an important reason for (initially)
restricting attention to generalizations of dynamics, rather than also admitting, say, 
nonlinear observables and state spaces more general than (projective) Hilbert spaces, 
is that it appears to be much more difficult to generalize the kinematics, while
still allowing for a consistent physical interpretation of such a generalized 
framework. A framework that incorporates standard quantum kinematics but in
which the dynamics is generalized, could then be regarded as a ``lowest order''
nontrivial approximation to a deeper nonlinear theory. But, in pursuing such
a strategy, it would be very useful to know if the standard kinematical framework
(together with certain other, physically reasonable assumptions) imposes any constraints on the nature of the dynamics.
It is the purpose of this note to critically re-examine two recent arguments
for the linearity of quantum dynamics in this regard. The first argument
is basically that relativistic causality requires quantum dynamics to be linear \cite{Simon1}.
It involves spatially separated, but entangled systems and is a different version 
of an older, essentially identical argument \cite{Wald,Gisin1}. The improvement 
with respect to the earlier version of the argument is argued to lie in the 
supposed facts that the projection postulate need not be assumed to establish
the argument and that a certain condition, known as ``complete positivity'', is implied by the properties of the dynamics.
It will be seen however that, although it may be possible to establish the 
argument without using the projection postulate, additional assumptions would
then have to be made. In any case, without the projection postulate, the
linearity of dynamics does not merely follow - as contended in Ref.
\cite{Simon1} - from quantum kinematics and relativistic causality\footnote{A somewhat similar criticism has been raised before 
\cite{Bona}.}. It will also be pointed out furthermore, that the remainder of 
the argument actually establishes a condition weaker than
complete positivity. The second recent argument for linear quantum dynamics mentioned 
above, is basically that if time-evolution is implemented by a well-defined map 
on the space of density operators, such that all possible ways of ``experimentally
realizing'' an application of the map to a given density operator always yield 
the same result, then this map must be linear \cite{Jordan2}. 
No entanglement is used in this argument and it does not suffer from a loophole that 
remains in the argument based on relativity. 
However, the statement that the assumed property of the map implies its linearity,
is almost a trivial assertion, whereas it will be argued that there are no good
reasons (apart from causality) to require that a map implementing time-evolution 
on a system should in fact satisfy this property.

\section{Entangled Systems : Linear Dynamics and Causality}\label{linearityargument1}
Consider a composite system, described in terms of the tensor product, $\Hilbert_{I} \otimes \Hilbert_{II}$, 
where the Hilbert spaces $\Hilbert_{I}$ and $\Hilbert_{II}$ are assumed to be finite-dimensional 
and are thought of as supporting states in causally disconnected regions of spacetime,
$\mathcal{O}_I$, $\mathcal{O}_{II}$. It is furthermore assumed that the total
system has been prepared in some pure entangled state, $|\Psi \rangle$, that
is taken to be normalized. Physical conditions in $\mathcal{O}_I$ are completely described by the reduced
density operator, $\rho_I = \mbox{Tr}_{II} \rho$, $\rho := |\Psi \rangle \langle \Psi|$,
but it is important to note that any statistical mixture associated with this
density operator is an improper mixture in the present context\footnote{It is recalled that any density
operator, $\rho$, has infinitely many \emph{$\rho$-ensembles} associated with
it, i.e. statistical mixtures, $\{ w_i , P_{\psi_i} \}$, such that
$\rho = \sum_i w_i P_{\psi_i}$, where $P_{\psi_i} = | \psi_i \rangle \langle \psi_i |$, 
$||\psi_i||=1$ (the states $|\psi_i \rangle$ are not necessarily orthogonal)
and $0 < w_i < 1$, $\sum_i w_i = 1$ \cite{Hughston} (in the pure case, all of
these ensembles are of course trivial; i.e. $P_{\psi_i} = P_{\psi}$, for
some fixed state $| \psi \rangle$, for all $i$).} \cite{dEspagnat1,*dEspagnat2}.
In other words, for any $\rho_I$-ensemble, $\{ w_i , P_{\psi_i} \}$, one cannot
consistently interpret $\rho_I$ as describing a large ensemble of
physical systems, such that a fraction $w_i$ of these systems are in the exact
quantum state $|\psi_i \rangle$ with certainty. 
Consequently, with this interpretation of $\rho_I$, the only object of physical relevance on which
an operator, \$, representing dynamical evolution in $\mathcal{O}_I$, can act,
is $\rho_I$ itself. In particular, it is physically meaningless to ask how 
the individual projectors, $P_{\psi_i}$, of a particular $\rho_I$-ensemble
are affected by the dynamics. The only consistency requirement that follows from
this is that if $\mathsf{T}$ denotes the operator implementing dynamical evolution on
the composite system (where both $\mathsf{T}$ and \$ are imagined to refer to the same
fixed time-interval in some particular reference frame), one must have $ \$ \rho_I = \mbox{Tr}_{II} \mathsf{T} \rho$.\\
It is an important mathematical fact, that for any $\rho_I$-ensemble,
$\{ w_i , P_{\psi_i} \}$, it is always possible (after enlargening the Hilbert
space $\Hilbert_{II}$ to another finite-dimensional Hilbert space, $\Hilbert_{II}'$, 
if necessary), to express the state $| \Psi \rangle$ as
\begin{equation}\label{Schmidt}
| \Psi \rangle \; =\; \sum_i \sqrt{w_i} |\psi_i \rangle | \phi_i \rangle
\end{equation}
where the $|\phi_i \rangle$ constitute an orthonormal basis for $\Hilbert_{II}$ \cite{Hughston}.
If the projection postulate is assumed, this physically means
that the $\rho_I$-ensemble $\{ w_i , P_{\psi_i} \}$ can be generated by local
operations in $\mathcal{O}_{II}$ (i.e. after first introducing an extra local
quantum system in some standard state, a so-called ancilla, if necessary, to
enlargen the original Hilbert space $\Hilbert_{II}$ \cite{Peres}, and then
perform a measurement in the basis $\{ | \phi_i \rangle \}$) and that this $\rho_I$-ensemble
now describes a statistical mixture which is proper (so that it is possible
\cite{dEspagnat1,*dEspagnat2} to consistently assign an ignorance interpretation
to $\{ w_i , P_{\psi_i} \}$). In other words, if the projection postulate is assumed, 
any (proper) $\rho_I$-ensemble in $\mathcal{O}_{I}$
can be generated by local measurements in $\mathcal{O}_{II}$. Now, let \$ 
again denote the operator implementing dynamical evolution in $\mathcal{O}_{I}$
and assume this operator to be defined initially on pure states only.
Since the time-interval associated with \$ is arbitrary, one must have
\begin{equation}\label{linear1}
\sum_i w_i \$ P_{\psi_i} \; = \; \sum_i \tilde{w}_i \$ P_{\tilde{\psi}_i}
\end{equation}
for any two $\rho_I$-ensembles $\{ w_i , P_{\psi_i} \}$ and $\{ \tilde{w}_i , P_{\tilde{\psi}_i} \}$, 
as otherwise it would be possible to communicate at superluminal speeds.
Since $\{ w_i , P_{\psi_i} \}$, $\{ \tilde{w}_i , P_{\tilde{\psi}_i} \}$ and $\rho_I$
are arbitrary, this means that \$ can be extended to a map, also denoted by
\$, on the space, $\Xi(\Hilbert_I)$, of density operators on $\Hilbert_I$,
which is linear in the sense that
\begin{equation}\label{lindyn}
\$ \rho_I \; = \; \$ \sum_i w_i P_{\psi_i} \; = \; \sum_i w_i \$ P_{\psi_i}
\end{equation}
for any $\rho_I \in \Xi(\Hilbert_I)$ and any $\rho_I$-ensemble $\{ w_i , P_{\psi_i} \}$
(alternatively, it could have been assumed that \$ is defined on $\Xi(\Hilbert_I)$
to begin with, so that the left-hand side of Eq. (\ref{lindyn}) represents the
time-evolved state in $\mathcal{O}_{I}$ in the case where no measurement has
been performed in $\mathcal{O}_{II}$; the fact that the state in $\mathcal{O}_{I}$
should be independent of the conditions in $\mathcal{O}_{II}$ then establishes Eq. (\ref{lindyn})).\\
Both earlier arguments for linear quantum dynamics \cite{Wald,Gisin1} have the 
above general structure of exploiting EPR-correlations to the extent that different 
descriptions of the same state in $\mathcal{O}_{I}$, that exist as a result of
what has happened in $\mathcal{O}_{II}$, must dynamically evolve into the same
state to avoid superluminal signalling. These arguments
assume the projection postulate, either implicitly or explicitly. In the more
recent version of the argument \cite{Simon1} this postulate is not assumed, but
other than that, the argument is identical to that given here. It will now be
seen however, that without the projection postulate, the argument is inconclusive. 
The crucial point of the previous discussion is that in order to establish the linearity of \$, 
it is necessary to assume that the act of carrying out a measurement in $\mathcal{O}_{II}$ induces physical
effects in $\mathcal{O}_{I}$, in such a way that the interpretation of a specific 
$\rho_I$-ensemble (associated with a particular orthonormal basis for $\Hilbert_{II}$, 
for a given total state $| \Psi \rangle$) undergoes a transition from an improper to a proper mixture 
(and it is rather clear that such a transition must correspond to something physical, as 
the two types of mixture refer to distinct experimental situations).
But, in terms of the mathematical formalism, it is not clear how to incorporate
this assumption without assuming that some appropriate form of the projection postulate holds
true, unless some specific further assumptions are made with regards to
\emph{interpreting} the mathematical formalism. 
To see this, the standard von Neumann account of measurements - which is assumed
explicitly in the argument of Ref. \cite{Simon1} - is briefly reviewed first, 
in order to determine the result of carrying out a measurement in $\mathcal{O}_{II}$. 
If the measurement apparatus is treated quantum mechanically, the result of an idealized 
measurement in $\mathcal{O}_{II}$ according to von Neumann is the state
\begin{equation}\label{measure1}
| \tilde{\Psi} \rangle \; := \; \sum_i \sqrt{w_i} | \psi_i \rangle | \phi_i \rangle | A_i \rangle
\end{equation}
where the apparatus states, $| A_i \rangle$, represent definite ``pointer readings''
corresponding to a measurement in the basis $|\phi_i \rangle $. 
Von Neumann's point was that if everything is to be treated in quantum mechanical
terms, one could insert arbitrarily long, finite chains of further physical
systems (i.e. other measurement apparatuses, sensory and nervous systems of 
``observers'', etc.), with which the original system plus measurement 
apparatus becomes entangled and that, in order to account for definite observational
facts, it ultimately does not matter exactly \emph{where} in this chain projections 
are introduced. 
Of course, the problem of how to account for definite observational facts within
quantum theory - i.e. the ``measurement paradox'' - and in particular, the issue
of whether it is necessary to introduce projections to resolve this paradox,
has been the source of much controversy and some remarks are in order here.
However, first note that from the viewpoint of the general discussion, the
assumption of von Neumann measurements is somewhat odd,
since such a notion of measurements entails the assumption of a linear - in 
fact unitary - dynamical evolution of the total system before and during the
measurement process. But the implication of considering the possibility of a
generalized dynamics is that there could be physical conditions for which this
notion of measurements is inadequate, in which case it is not clear that the
argument for linear dynamics does not break down. However, this difficulty
can be circumvented simply by assuming the projection postulate, without making
any assumptions about the actual measurement interactions.
But if this point is ignored, the problem from the viewpoint of the general
discussion is whether proper mixtures can be generated in $\mathcal{O}_{I}$,
as a result of carrying out von Neumann measurements in $\mathcal{O}_{II}$.
Now, the logical inference of von Neumann's account of measurements - and it is
again emphasized that this account is based on the assumption that the dynamics
of the total system remains unitary, as the degrees of freedom of the 
system of interest become entangled with increasingly more degrees of freedom
of other systems in the chain - strongly seems to be that \emph{if} it is believed 
necessary to introduce projections to account for definite observational facts, 
the only reasonable place to do so would be the point
where the ``consciousness'' of an ``observer'' enters. Although such an idea has
indeed been taken seriously by a number of physicists, most notably von Neumann
himself and Wigner, it is easily seen to lead to a number of serious difficulties.
An immediate mathematical problem is the fact that the argument \emph{assumes} 
from the outset that only ``classically reasonable'' states are ultimately allowed 
to occur or be ``experienced'', or, equivalently, it assumes that there are 
preferred bases in Hilbert space\footnote{The Schmidt polar decomposition - 
given by Eq. (\ref{Schmidt}) in the case where the $w_i$ and $| \psi_i \rangle$ 
are respectively the nonzero eigenvalues and eigenstates of $\rho_I$ - does not 
correspond to a preferred basis for $\Hilbert_I \otimes \Hilbert_{II}$
in this regard, since the eigenstates of $\rho_I$ are not in general ``classically
reasonable''. However, the idea of ``environment induced decoherence'' leads, under
certain assumptions, to a preferred basis of apparatus states, $| A_i \rangle$,
a so-called ``pointer basis''. See the main text for further discussion.},
something which has no natural place in the mathematical formalism of quantum
theory and therefore needs further justification.
Far more problematic however, are the conceptual implications of the view that
it is the act of conscious observation that reduces the quantum state.
As has been stressed by a number of authors (see e.g. Ref. \cite{Bell1,*Bell2}), 
it is far from clear what ``consciousness'' precisely is and what amount of it would furthermore
be required to effect state reduction. In addition, it has been pointed out
that the view ``does not respect the symmetry that the facts are invariant
under interchange of observers'' \cite{ConwayKochen}, constituting the so-called 
concordance problem. In other words, the upshot of following von Neumann's
account of measurements all the way, is that it leads to a view very similar to
solipsism.\\ 
The reason for reviewing von Neumann's account of the measurement process - of
which projections are a part - is that it is easy to shift from the solipsist view, in which the
quantum state is ``objectively'' reduced by a single observer's consciousness, 
to the opposite extreme, in which all alternatives continue to exist in parallel and
the state reduction only takes places subjectively - in the mind of the observer.
That is, it seems that if one follows von Neumann's account of measurements all 
the way, one may just as well dispose of projections altogether and adopt an
essentially Many Worlds type of perspective\footnote{This characterization of
the Many Worlds viewpoint is intended to be flexible. It may either refer to
a literal perspective, with ``copies'' of a single ``observer''
in different states of perception co-existing in different worlds, or it may
refer to a ``Many Minds'' perspective, in which there is only a single world,
but with a ``branching of minds'', reflecting the different states of perception.}.
Is this what is implicitly assumed in the argument of Ref. \cite{Simon1} ? Although it 
strongly appears that such a view is logically implied by employing von Neumann's
account of measurements without using projections, it may be objected that there
are a number of ``no-collapse'' interpretations \cite{Bub}, which differ with
respect to specific details. But the argument of Ref. \cite{Simon1} does not
even mention any particular such interpretation and it is far from clear that
the argument is actually interpretation independent - within the class of such
no-collapse interpretations. Furthermore, contrary to what has sometimes been
claimed in the literature (see e.g. Ref. \cite{Zurek3}), decoherence is actually
of no help as far as generating definite observational facts is concerned.
For what the theory of ``environment induced decoherence''
\cite{Zurek1,*Zurek2,*JoosZeh} merely shows, is that upon modeling the external
environment in an appropriate way and inserting it in the von Neumann chain,
quantum interference effects quickly wash out, as a result of the system plus
measurement apparatus interacting with the external environment.
More precisely, what the theory shows is that, under appropriate conditions,
the reduced density operator, $\rho_{I,II,M_{II}}$, associated with
the system and measurement apparatus, rapidly becomes diagonal with respect
to the specific ``pointer basis'' of apparatus states, $|A_i \rangle$, i.e.
$\rho_{I,II,M_{II}} = \sum_i w_i P_{\psi_i} \otimes P_{\phi_i} \otimes P_{A_i}$.
But this only shows that $\rho_{I,II,M_{II}}$ \emph{behaves} as a
probabilistic mixture. To say that a particular probabilistic mixture is actually
realized as a result of the measurement process - as necessary for the linearity
argument - requires adopting a specific view on how the quantum state at the
Hilbert space level, which is still a superposition of macroscopically distinct
states, i.e. Eq. (\ref{measure1}), is to be interpreted\footnote{For further
discussion of this point, see e.g. Ref. \cite{Adler,*BassiGhirardi1}.}.
If one insists on not using the projection postulate, there are several possible
views that can be taken, as mentioned above. One may then attempt to establish
the argument for linear dynamics within a specific such view, but the validity
of the argument would then depend on the validity of the specific view adopted.
For instance, in the case of a Many Worlds type of interpretation, it is not clear
that outcomes obeying the square modulus rule can be shown to arise as ``typical''
by postulating an appropriate measure on the world Hilbert space. Even when
decoherence effects are taken into account, the state represented by Eq. (\ref{measure1}) can still
be decomposed in infinitely many \emph{different} ways and these must all have
measure zero\footnote{For critical reviews of the Many Worlds programme in general, 
see Ref. \cite{Kent4} and further references therein. For critical discussions concerning 
the status of proposed derivations of the square modulus rule within this programme, see Ref. \cite{Stein,*Squires1,*Barnumetal,*SchlosshauerFine}.}.
More generally, if one insists on an entirely linear, orthodox quantum treatment of all physical
systems, it is necessary to explain how the environment enters a sufficiently
``classical state'', necessary to trigger decoherence in the first place.
From a more fundamental viewpoint, this merely shifts the whole problem of
\emph{even accounting for the appearance of classical behaviour}.
It is of course true that interpretational models in which state reduction is
viewed as a real physical phenomenon also encounter some serious difficulties. However, this
does not affect the argument for linear dynamics, as projections may simply be
introduced on an ad hoc basis - although they are of course motivated by the idea
that state reduction is objective. Furthermore, as already pointed out, there
is something intrinsically ill-motivated about the entire argument if the standard
von Neumann account of measurements is assumed to apply universally.\\
The upshot of the previous discussion is that the argument for linear dynamics,
as presented in this section, is not conclusive if the projection postulate
is not assumed. On the other hand, within the von Neumann context of measurements,
it is clear from Eq. (\ref{measure1}) that if the apparatus was observed in
the particular state $|A_k \rangle$, say, the state of the system in $\mathcal{O}_{I}$
is $| \psi_k \rangle$, if it is assumed that immediately after the measurement 
was performed, the state $|\tilde{\Psi} \rangle$ is reduced to 
$(\mathbbm{1} \otimes \mathbbm{1} \otimes P_{A_k})|\tilde{\Psi} \rangle = \sqrt{w_k}| \psi_k \rangle | \phi_k \rangle | A_k \rangle $.
This option is explicitly rejected in Ref. \cite{Simon2}, because it is pointed
out that it could happen that the system in $\mathcal{O}_{II}$ is destroyed by the measurement, something
which would effectively modify the Hilbert space of the system according to 
$\Hilbert_I \otimes \Hilbert_{II} \otimes \Hilbert_M \rightarrow \Hilbert_I \otimes \Hilbert_M$
(with $\Hilbert_M$ denoting the Hilbert space of the measurement apparatus).
Instead of Eq. (\ref{measure1}) one would then have
\begin{equation}\label{measure2}
| \Psi  \rangle \; \longrightarrow \; | \mathring{\Psi}  \rangle \; := \; \sum_i \sqrt{w_i} | \psi_i \rangle | A_i \rangle 
\end{equation}
However, in this particular case, even though one cannot use the projection 
postulate on $\Hilbert_I \otimes \Hilbert_{II}$ (as there is no projection operator 
from $\Hilbert_I \otimes \Hilbert_{II}$ to $\Hilbert_I$), it is still possible
to use the postulate on $\Hilbert_I \otimes \Hilbert_M$, or a larger Hilbert space.
Thus, by assuming (some appropriate form of) the projection postulate, it becomes
possible to generate proper mixtures at a distance and the linearity of dynamics
can be established. According to the argument in Ref. \cite{Simon1}, it is merely
the trace rule that ``leads to the preparation at a distance of probabilistic mixtures''.
More precisely, it is argued that the state in $\mathcal{O}_{I}$ can be deduced 
by the observer in $\mathcal{O}_{II}$, by calculating the conditional probability 
for every projection operator for the system in $\mathcal{O}_{I}$, given a definite 
measurement outcome in $\mathcal{O}_{II}$.
If it is again supposed that the total state is given by Eq. (\ref{Schmidt}) 
and that the observer in $\mathcal{O}_{II}$ carries out a measurement in the orthonormal basis $\{ | \phi_i \rangle \}$,
the probability to find the system in $\mathcal{O}_{I}$ in the state $P_{\psi_j}$,
conditional upon the system in $\mathcal{O}_{II}$ having been found in the
state $P_{\phi_i}$, is just $\mbox{Tr} \rho (P_{\psi_j} \otimes P_{\phi_i}) / \mbox{Tr} \rho ( \mathbbm{1} \otimes P_{\phi_i}) = \mbox{Tr} P_{\psi_i} P_{\psi_j}$
and according to Ref. \cite{Simon1}, the state of the first system in the present
context is given by $\mbox{Tr}_{II} \rho (\mathbbm{1} \otimes P_{\phi_i}) / \mbox{Tr} \rho ( \mathbbm{1} \otimes P_{\phi_i}) = P_{\psi_i}$.
Thus, what this argument actually appears to amount to, is just the usual 
plausibility argument for the projection postulate; if a measurement of
$P_{\psi_i}$ on the $\mathcal{O}_{I}$-system is certain to give the eigenvalue
$1$, then the state of this system actually \emph{is} the eigenstate $P_{\psi_i}$. 
In other words, the authors of Ref. \cite{Simon1} appear to claim that they are 
able to \emph{derive} the projection postulate from the kinematical framework of 
quantum theory. But, in view of the existence of no-collapse interpretations, such 
a claim would clearly be contentious. 

\section{Further Discussion}\label{discussion}
To summarize the discussion of the previous section, it was argued - in concordance
with earlier arguments - that the absence of superluminal signalling, i.e. relativistic causality, together with the standard
kinematical framework for quantum theory and (some appropriate form of) the
projection postulate, imply that the dynamics of density operators is necessarily
linear and that, furthermore, in spite of what has been claimed in the literature,
the projection postulate assumption has so far not been demonstrated to be dispensible
for this argument. In this section, we address two types of question which 
are naturally raised by the above argument :
\begin{enumerate}
\item[(i)] Is the argument rigorous ? If loopholes remain, how plausible are they ?
\item[(ii)] Given that dynamical evolution on the space of density operators
is linear, what does this imply for the nature of dynamical evolution on the
corresponding Hilbert space ? What does it imply for the nature of dynamics
if the corresponding Hilbert space is a factor in a tensor product ?
\end{enumerate}
Concerning the first type of question, it is not difficult to see that the
paradoxical nature of preparing ensembles at a distance points to a loophole 
in the argument \cite{Jordan2,Kent1,Jordan}.
Given any subset, $S$, of Minkowski spacetime, let $I^+(S)$, $J^+(S)$, denote
respectively the ``chronological future'' and the ``causal future'' of $S$ (i.e.
$I^+(S)$, resp. $J^+(S)$, is the set of all points in Minkowski spacetime that
can be joined to a point in $S$ by a future directed curve, which is timelike,
resp. null or timelike, and which starts in $S$; cf. Ref. \cite{Wald2}). Let
$y$ furthermore denote the event in $\mathcal{O}_{II}$ at which a measurement
is carried out on the second system. The loophole is that the state of the first
system reduces \emph{causally} as a result of this measurement. That is, if the
second sytem was found in the state $P_{\phi_j}$, the state of the first system 
changes from $\rho_I$ to $\mbox{Tr}_{II} \rho (\mathbbm{1} \otimes P_{\phi_j}) / \mbox{Tr} \rho ( \mathbbm{1} \otimes P_{\phi_j}) = P_{\psi_j}$,
as the worldline in $I^+( \mathcal{O}_I)$ of the first system enters $J^+(y)$ 
(in Ref. \cite{Kent1} the notion of a ``local state'' of a system is introduced,
in terms of a limiting procedure of a sequence of spacelike hypersurfaces that tend 
to the past lightcone of the system's instantaneous spacetime location, thereby allowing 
the system's state to change in a causal manner, as a result of performing measurements on 
a second system with which the system is entangled).
Upon extending this construction to ensembles of entangled systems, it is thus
seen that proper mixtures are generated in a causal manner in $I^+(\mathcal{O}_I) \cap J^+(\mathcal{O}_{II})$,
as a result of performing measurements in $\mathcal{O}_{II}$. But it is clear
that there is now no linearity restriction on the dynamical evolution of the
first system. Thus, if state reduction is assumed to occur in a causal fashion, 
a nonlinear dynamics of density operators is consistent with causality. It is
possible to address this view on state reduction by experiment. If $\Delta$
denotes the minimum invariant distance between $\mathcal{O}_I$ and $\mathcal{O}_{II}$,
it takes at least a time $\Delta$, according to this view, for the mixture in
$I^+(\mathcal{O}_I)$ to become proper. Thus, by letting the experimenters in
$\mathcal{O}_I$ and $\mathcal{O}_{II}$ make prior arrangements on what measurements
to perform and by choosing $\Delta$ large enough, it is possible for the experimenters
to determine - after comparing their measurement results - whether the mixture
in $\mathcal{O}_I$ was indeed improper for at least a time $\Delta$. In order 
to definitely close this loophole in the argument, it is estimated that
$\Delta$ should be of the order of $3 \, 10^4 \, \mbox{km} \: \simeq \: 0.1 \, \mbox{lightsec}$ \cite{Kent3}.
However, although it seems to be consistent with present experimental facts
to think that state reduction occurs in a causal fashion, it is difficult to
believe that this corresponds to the actual state of affairs in Nature, as it
would entail a radical departure from what is currently seen in Bell-type experiments
as soon as the measurement events in such experiments can definitely be regarded
as spacelike separated\footnote{The ``collapse locality loophole'' discussed
in the text should not be confused with the so-called ``locality loophole''. The
latter loophole depends on the fact that it is difficult to experimentally arrange
that the events at which the measurement-settings are (randomly and independently)
chosen are at spacelike separation from each other and from the source event
and it seems to be generally accepted at present that this loophole has effectively been closed
\cite{Aspectetal,*Weihsetal}. It is also important to note that the collapse 
locality loophole (in spite of its name) applies equally well to no-collapse 
interpretations, as may be inferred from the discussion of measurements
in the preceding section (the crucial point is whether the act of ``property
determination'', as defined by the measurement events, occurs in spacelike
separated regions).}. The problem with this strategy more generally, 
is that it will never be possible to close \emph{all} loopholes completely
in Bell-type experiments to everyone's satisfaction\footnote{For instance, if one denies the freedom of experimenters to choose the orientation
of their measurement apparatuses, the nonlocality demonstrated in Bell-type
experiments can be ``explained'' by arguing that Nature is in an unmitigated
conspiracy to consistently give the appearance to behave nonlocally in certain
contexts, by denying experimenters the freedom to prove otherwise. But, as has
been noted elsewhere \cite{ConwayKochen}, it is very difficult to take science
seriously upon such a conception of Nature.} (see also the remarks below).
Nevertheless, with respect to 
the specific loophole discussed here, it is reasonable to require that experiments 
address it in due time (and to moreover expect that, when this is done, the 
loophole will be closed).\\
The second type of question listed above concerns the ramifications of the linearity 
of time-evolution of (reduced) density operators.
It is clear that a linear time-evolution on a Hilbert space, \Hilbert, implies
a linear time-evolution on the corresponding space of density operators, $\Xi(\Hilbert)$.
Conversely however, a linear time-evolution on $\Xi(\Hilbert)$ - again denoted 
by \$ - in itself implies very little for the nature of time-evolution at the
Hilbert space level. If \$ is invertible, it maps pure states into pure states
and it is not hard to show that this implies that \$ preserves the absolute
values of inner products \cite{Jordan2}. Hence, by Wigner's theorem, the action
of \$ then takes the form
\begin{equation} 
\$ \rho \; = \; U \rho U^{\dagger}
\end{equation}
with $U : \Hilbert \rightarrow \Hilbert$ unitary (if \$ is Markovian, the possibility
of an anti-unitary map is excluded).\\
In Ref. \cite{Simon1} it is argued that the linearity of \$, together with the
way it is defined (i.e. as a dynamical map from density operators to density operators),
implies its complete positivity\footnote{A general linear operator, $T : L(\Hilbert) \rightarrow L(\Hilbert)$,
with $L(\Hilbert)$ denoting the linear space of bounded linear operators on \Hilbert,
is said to be ``completely positive'', if $T \otimes \mathbbm{1}$ preserves positivity 
on $L(\Hilbert \otimes \Hilbert ')$, for any finite-dimensional Hilbert space 
$\Hilbert '$ (i.e. if $T \otimes \mathbbm{1}$ applied to any bounded, non-negative 
linear operator on $\Hilbert \otimes \Hilbert '$ again yields a bounded, non-negative 
linear operator, for any finite-dimensional Hilbert space $\Hilbert '$).
The definition given in Refs. \cite{Preskill,BreuerPetruccione} is more general
than the present definition (in the sense that $\Hilbert '$ is not restricted
to be finite-dimensional and the operators in question are not required to be
bounded), whereas the definition in terms of $C^*$-algebras given in Ref. \cite{Davies} 
is slightly more abstract, but is in fact easily seen to include the definition 
given here. It is then possible to prove
\cite{Davies} that a normal, non-negative linear operator, $T$, on $L(\Hilbert)$
is completely positive if and only if it has an operator-sum representation,
i.e. if and only if it can be written in the form $T(X) = \sum_i M^{\dagger}_i X M_i$,
for some set $M_i \in L(\Hilbert)$ and for all $X \in L(\Hilbert)$ (normality
of $T$ basically means that a certain convergence condition on the subspace of
self-adjoint operators is preserved by $T$). The operator-sum representation
of a (Markovian) superoperator is used to derive the infinitesimal form of the
dynamics, as given by Eq. (\ref{Lindblad}).}.
\$ \nolinebreak would then be a superoperator, which means that it can be made to correspond -
via its operator-sum representation - to a unitary mapping in a larger Hilbert
space \cite{Preskill,BreuerPetruccione}. As will now be shown however, the argument 
given in Ref. \cite{Simon1} actually only establishes a condition weaker than complete positivity.
To this end, consider a system described by a Hilbert space \Hilbert and with
time-evolution determined by \$, which is entangled with another system, 
described by a Hilbert space $\Hilbert '$ and with no time-evolution.
It is then argued in Ref. \cite{Simon1} that the time-evolution of the composite system is given by 
$\$ \otimes \mathbbm{1}$ and that, furthermore, the fact that this map should take any density
operator of the composite system into another such operator ``is exactly
the definition of complete positivity for the map \$ \cite{Preskill}''. But
the definition of complete positivity given in Refs. \cite{Preskill,BreuerPetruccione} requires
\emph{every} non-negative operator on $\Hilbert \otimes \Hilbert '$ to be mapped
into another such operator. Without additional information, the argument 
of Ref. \cite{Simon1} therefore does not establish that \$ is necessarily 
completely positive. However, an initially more serious difficulty with the argument 
is the fact that time-evolution of the composite system is taken to be represented
by $\$ \otimes \mathbbm{1}$. The operator \$ represents dynamical evolution on $\Xi(\Hilbert)$
and the operator $\$ \otimes \mathbbm{1}$ thus naturally represents dynamical
evolution on $\Xi(\Hilbert \otimes \Hilbert ')$ for states of the form $\rho \otimes \rho '$,
$\rho \in \Xi(\Hilbert)$, $\rho' \in \Xi(\Hilbert ')$. A general state $\sigma \in \Xi(\Hilbert \otimes \Hilbert ')$
is not of this form however and it is therefore not a priori clear that $\$ \otimes \mathbbm{1}$
is well-defined on all of $\Xi(\Hilbert \otimes \Hilbert ')$. To this end, it
may be noted that a general state, $\sigma$, can be written as a sum of products
of operators \cite{BreuerPetruccione}, i.e. $\sigma = \sum_i \xi_i \otimes \chi_i$,
for certain sets of operators $\xi_i$, $\chi_i$ on \Hilbert and $\Hilbert '$,
respectively. However, it is not difficult to see that the $\xi_i$ cannot
all be simultaneously self-adjoint, non-negative and not traceless for every
Hilbert space $\Hilbert'$ (if this were the case, taking $\Hilbert ' = \Hilbert$,
it would by symmetry also have to be true for the $\chi_i$, but then
$\sigma$ would not be a general state, as any $\sigma$-ensemble is a proper mixture).
Thus, expressing $\sigma$ as a sum of products does not make it clear that the
action of $\$ \otimes \mathbbm{1}$ is well-defined on $\sigma$. However, the
fact that all $\sigma$-ensembles are proper mixtures, implies that it is sufficient 
to determine whether the action of $\$ \otimes \mathbbm{1}$ is well-defined on a 
general pure state $| \Psi \rangle \in \Hilbert \otimes \Hilbert '$. In terms 
of the Schmidt polar decomposition for $| \Psi \rangle$, i.e. $| \Psi \rangle = \sum_i \sqrt{w_i} | \varrho_i \rangle | \varphi_i \rangle$,
with both sets $| \varrho_i \rangle$, $| \varphi_i \rangle$ orthonormal, this
action would be given by
\begin{equation}\label{compdyn}
\sum_{i,j} \sqrt{w_i w_j} | \varrho_i \rangle \langle \varrho_j | \otimes | \varphi_i \rangle \langle \varphi_j |  
\longrightarrow \sum_{i,j} \sqrt{w_i w_j} \$ \bigl( | \varrho_i \rangle \langle \varrho_j | \bigr) \otimes | \varphi_i \rangle \langle \varphi_j |
\end{equation}
But it is not clear that the right-hand side of this expression is well-defined,
since the operators $| \varrho_i \rangle \langle \varrho_j |$ are not elements
of $\Xi (\Hilbert)$ for $i \neq j$ (if a general expansion for $| \Psi \rangle$
is used, the same difficulty is encountered). It is of course conceivable that
the domain of definition of \$ naturally extends to a larger set, which
includes all operators of the form $| \varrho_i \rangle \langle \varrho_j |$
(e.g. the set of all bounded linear operators on \Hilbert). For instance, if
\$ is completely positive, it is clear that the right-hand side of Eq. (\ref{compdyn})
is formally well-defined, by expressing \$ in terms of its operator-sum representation.
But it is not necessary for \$ to have an operator-sum representation in order
for the right-hand side of Eq. (\ref{compdyn}) to be formally well-defined.
For instance, the map $\$_T$, which takes every density operator into its transpose,
implements a two-step cyclic time-evolution on $\Xi (\Hilbert)$ and is such that
$\$_T \otimes \mathbbm{1}$ is well-defined on all of $\Xi (\Hilbert \otimes \Hilbert ')$
(via the right-hand side of Eq. (\ref{compdyn})). But the map $\$_T$ is not completely
positive \cite{Preskill} and $\$_T \otimes \mathbbm{1}$ does not map $\Xi (\Hilbert \otimes \Hilbert ')$
into itself. However, the point here is that it is conceivable that \$ is such that
its domain of definition can be naturally extended to include all operators of
the form $| \varrho_i \rangle \langle \varrho_j |$ \emph{and} that $\$ \otimes \mathbbm{1}$
maps $\Xi (\Hilbert \otimes \Hilbert ')$ into itself for any finite-dimensional Hilbert space $\Hilbert '$,
but that \$ is nevertheless not completely positive in the technical sense of
the definition, so that it does not have an operator-sum representation.
There would then have to be at least one finite-dimensional Hilbert space, $\Hilbert '$, and a bounded,
non-negative operator on $\Hilbert \otimes \Hilbert'$, that is not self-adjoint,
and which is mapped by $\$ \otimes \mathbbm{1}$ into an operator on $\Hilbert \otimes \Hilbert'$, 
that is either not bounded, not non-negative, or both.
As far as the present author is able to tell, no mathematical theorem rules out this 
possibility, although he knows of no example of a map with these properties.\\
As noted above, if \$ is completely positive, it can be made to correspond to a
unitary mapping in a larger Hilbert space \cite{Preskill}. This property of \$ 
however does not constrain dynamics to be unitary - or even linear -
on the original Hilbert space $\Hilbert_{I} \otimes \Hilbert_{II}$ that was used
to prove the linearity of \$. For instance, on using again the Schmidt polar 
decomposition for a general state $| \Psi \rangle \in \Hilbert_{I} \otimes \Hilbert_{II}$,
the following evolution for $| \Psi \rangle := | \Psi (0) \rangle$ 
\begin{equation}
| \Psi (0) \rangle \; := \; \sum_i \sqrt{w_i} | \varrho_i \rangle | \varphi_i \rangle \; \longrightarrow \; \sum_i \sqrt{w_i} e^{i w_i \theta t} | \varrho_i \rangle | \varphi_i \rangle \; =: \; | \Psi (t) \rangle   \qquad \quad \theta \in 
\mathbb{R} \backslash \{ 0 \}
\end{equation}
is clearly nonlinear, even though the reduced dynamics is linear (and in fact
trivial\footnote{This example is taken from Ref. \cite{Kent5}. For a different example, see Ref. \cite{Ferrero}.}).
Furthermore, it was recently demonstrated \cite{JoShaSu} that a unitary dynamics
on a composite Hilbert space in general does not correspond to a reduced dynamics
that is completely positive. Since a composite unitary dynamics clearly implies
a reduced dynamics of density operators that is linear, this thus confirms the 
above claim that the linearity of \$ (together with the way it is defined) does not imply its complete positivity.\\
Linearity and complete positivity of \$ imply that it is (infinitesimally) 
given by the Lindblad form\footnote{Strictly speaking it is also necessary to impose
the condition that \$ be of a Markovian nature. This has been assumed (implicitly)
throughout the discussion.} \cite{Lindblad}
\begin{equation}\label{Lindblad}
\dot{\rho} \; = \; -i [H, \rho] \: + \: \sum_j \bigl( L_j \rho L^{\dagger}_j  \: - \: \frac{1}{2} \{ \rho , L^{\dagger}_j L_j \} \bigl)
\end{equation}
where $H$ represents the Hamiltonian of the system, $\{ \cdot , \cdot \}$
denotes the anticommutator and the $L_j$ are so-called quantum jump operators.
At the Hilbert space level, the class of evolution equations defined by Eq. (\ref{Lindblad})
includes the nonlinear, stochastic modifications of the Schr\"odinger equation,
that were introduced to solve the measurement paradox \cite{Gisin1,Pearle1,*Pearle2,GRW,*GPR,Diosi,Percival}
(for a recent review, see Ref. \cite{BassiGhirardi}).
For nonzero $L_j$, Eq. (\ref{Lindblad}) describes the loss of quantum coherence,
i.e. it describes the evolution of pure states into mixed states. Apart from
possible applications to the measurement paradox, such evolution has been argued
to be relevant to describe the process of black hole evaporation \cite{Wald,Hawking,Hartle1,*Hartle2} 
(for an astrophysical black hole such evaporation should occur as soon as its
Hawking temperature drops below the temperature of the cosmic background radiation).
The loss of quantum coherence is often regarded as something unphysical, since
it is expected to give rise to violations of either causality or energy-momentum
conservation \cite{BaSuPe}. It has been shown however \cite{UnruhWald} that it
is possible to define a class of Markovian models in which pure states evolve into mixed
states, but which do not lead to pathological behaviour at scales accessible
to ordinary laboratory physics. In addition, the argument of Ref. \cite{BaSuPe}
is based on the assumption that the process of black hole formation and evaporation
is effectively modeled by the evolution (\ref{Lindblad}), whereas it is unclear
that the Markovian character of this evolution is physically reasonable
in the case of black holes \cite{UnruhWald}.\\
Finally, it is important to note that the argument for linear dynamics does
not continue to be valid in contexts that go beyond standard quantum kinematics.
For instance, Weinberg \cite{Weinberg1,*Weinberg2,*Weinberg3} has proposed
a nonlinear generalization of quantum theory, in which observables are represented
by real-valued functions on \Hilbert, homogeneous of degree one in $\Psi$, $\overline{\Psi} \in \Hilbert$
(where $\Psi$ and $\overline{\Psi}$ are taken as independent variables). Within
this framework, the observables of standard quantum theory are represented by 
the subset of bilinear, real-valued functions - corresponding to expectation 
value functions of self-adjoint linear operators. This naturally suggests a
generalization of the notion of a state as a general probability distribution,
$\rho$, on ``quantum phase space'', i.e. the projective Hilbert space, \PHilbert
(where it is recalled that \PHilbert is equipped with a symplectic structure
that is generated by the imaginary part of the inner product on \Hilbert). The
trace rule for the expectation value of a standard quantum observable, which 
generalizes to the value of a general Weinberg-observable for pure states, 
then naturally extends to a $\rho$-weighted integral over \PHilbert of a Weinberg-observable
for mixed states. On such an account, the density operator, $\hat{\rho}$, of ordinary quantum
theory is only the first moment of a one-dimensional projection operator in the
state $\rho$ (i.e. the $\rho$-weighted integral of a one-dimensional projection operator 
over \PHilbert), which implies that $\hat{\rho}$ does not in 
general contain all the information about a system, when dealing with nonlinear
observables \cite{BrodyHughston1,*BrodyHughston2}. Failure to take this fact 
into account can lead to contradictory results \cite{Gisin2,*Polchinski}, although
it is not clear that a proper treatment of the quantum state guarantees the absence
of such results.

\section{Unentangled Systems : Linear Dynamics and Preservation of Mixture Independence}
As already remarked in the introduction, a different argument for the linearity
of quantum dynamics was recently advanced by Jordan \cite{Jordan2}. He considers
the dynamics for a system which is part of a larger system, but only so in a 
classical sense. That is, the system, which is denoted by S in Ref. \cite{Jordan2}, 
is combined with another system, denoted by R, but there is no entanglement between 
S and R. As will become clear in the following,
these specific assumptions are somewhat unfortunate, as they tend to deflect
attention away from the core of the argument. In order to bring it into accordance 
with the notation and language used in the previous two sections, Jordan's argument 
is first reformulated and slightly generalized.\\
Given any density operator, $\rho$, for S and any $\rho$-ensemble, $\{ w_i,P_{\psi_i} \}$,
Jordan considers two possible density operators for the combined system that
both have $\rho$ as a reduced density operator
\begin{equation}\label{Jordanstate}
\overline{\Pi}_{\{ w , P_{\psi} \}} \; := \;  \sum_i w_i P_{\psi_i} \otimes P_{\alpha_i}   \qquad \qquad  \Pi_{\rho} \; := \;  \rho \otimes \sum_i s_i P_{\alpha_i}
\end{equation}
with $P_{\alpha_i} := | \alpha_i \rangle \langle \alpha_i |$ time-independent,
orthogonal projectors for R, $0< s_i <1$ and $\sum_i s_i = 1$ (cf. Eqs. (2.4),
(2.7) respectively in Ref. \cite{Jordan2}). The interpretation of these density 
operators will be addressed shortly, something which is in fact crucial for the
entire argument. The key assumption made by Jordan is that the dynamics of S
does not depend on how S is ``embedded'' within the larger system. He also assumes
that the operator implementing time-evolution over some fixed time-interval
is defined both on $\rho$ directly and on the components $P_{\psi_i}$ of the
particular decomposition $\{ w_i , P_{\psi_i} \}$ of $\rho$. On denoting this
operator again by \$, it then immediately follows from Jordan's key assumption
that
\begin{equation}\label{lindyn2}
\$ \rho \; = \; \$ \sum_i w_i P_{\psi_i} \; = \; \sum_i w_i \$ P_{\psi_i}
\end{equation}
so that \$ is indeed linear.\\ 
The two above assumptions will now be questioned, by addressing the physical interpretation
of the two density operators $\overline{\Pi}$ and $\Pi$.
As for $\overline{\Pi}$, measurements on R that distinguish between the
orthonormal states $| \alpha_i \rangle$, generate the $\rho$-ensemble $\{ w_i,P_{\psi_i} \}$.
If the same measurement procedures for R are applied to $\Pi$ however, it seems
at first that the most one can say for S is that its state is given by
$\rho$, regardless of which state the system R is found to be in.
But since there is no entanglement, all $\rho$-ensembles are proper mixtures, 
and it is therefore possible to consistently think of the system S as actually being in some definite
state in $\Hilbert_S$ with a certain classical probability. Hence, in the situation
where the two states are prepared experimentally, in the case of $\Pi$, a fraction
$\tilde{w}_i s_j$ of a large number of combined systems is definitely prepared
in the state $P_{\tilde{\psi}_i} \otimes P_{\alpha_j}$, with $\{ \tilde{w}_i , P_{\tilde{\psi}_i} \}$
denoting any $\rho$-ensemble, whereas in the case of $\overline{\Pi}$, a fraction
$w_i$ of the combined systems is prepared in the state $P_{\psi_i} \otimes P_{\alpha_i}$.
Upon an ignorance interpretation of $\overline{\Pi}$ and $\Pi$, the difference
that remains between the two states is that there is no one-to-one correspondence 
between physically realized states in $\Hilbert_S$ and $\Hilbert_R$ in the latter case. 
With the ensembles realized physically, what evolves with time are the individual
pure states, $P_{\psi_i}$, $P_{\tilde{\psi}_i}$, so that \$ is initially only defined 
on the set of all such states. The time-evolved ensembles are then represented by
\begin{equation}
\sum_i w_i \$ P_{\psi_i} \otimes P_{\alpha_i}  \qquad \mbox{and} \qquad  \sum_{i,j} \tilde{w}_i s_j  \$ P_{\tilde{\psi}_i} \otimes P_{\alpha_j}
\end{equation}
for respectively $\overline{\Pi}$ and $\Pi$. Jordan's key assumption is that
these two different evolved states for the combined system have to result in the
same state for S, which immediately implies the equation analogous to Eq. (\ref{linear1})
for the current situation. In other words, the possibility of different
statistical mixtures corresponding to the same density operator evolving into
statistical mixtures corresponding to different density operators is ruled out
by assumption. As before, this assumption immediately implies that \$ can be
extended to a well-defined map on $\Xi(\Hilbert_S)$, which is linear in the
sense of Eq. (\ref{lindyn2}).\\
Thus, to summarize, because there is no entanglement, the assumption that \$
has to act directly on $\rho$ - as apparently suggested by the form (\ref{Jordanstate})
for $\Pi$ - is found to be unjustified. In a concrete experimental situation,
a particular $\rho$-ensemble is realized physically and it are the pure states
in such an ensemble on which \$ acts.\\
Let us now address Jordan's key assumption. At first, it appears to be
very reasonable on physical grounds, as there is no interaction between the systems
S and R. On closer inspection however, the issue of whether the assumption is
physically reasonable or not, has nothing to do with the absence of interactions 
between S and R, but instead turns out to crucially depend on a peculiar property
of density operators in ordinary quantum theory. An experimenter confronted with
a large ensemble of physical systems of which a fraction $w_i$ was prepared
in the pure state $| \psi_i \rangle$, cannot ascertain this fact according
to ordinary quantum theory. All he can know by doing experiments on the ensemble is the density operator,
$\rho$, determined by this particular ensemble. Thus, in standard quantum theory
it is possible that ensembles of systems, which were experimentally prepared 
\emph{differently}, are represented by the \emph{same} physical state. In fact,
what was said before shows that the statement that density operators are subject
to a linear dynamics is \emph{equivalent} to requiring that this property of
ensemble-independence of states is preserved by the dynamics. But, is it physically reasonable to impose
such a requirement ? From experience with ordinary quantum theory one is tempted
to think that it is, but if one is to \emph{derive} the linearity of quantum dynamics,
it is necessary to give a different motivation. In fact, it does not seem that
there is a good physical justification - apart from causality and expectations
based on experience with ordinary quantum theory - for requiring
that differently prepared ensembles corresponding to the same density operator 
always evolve into ensembles that again correspond to a single density
operator. Indeed, with a dynamics that physically distinguishes different mixtures belonging
to the same initial density operator, it becomes necessary to sharpen the interpretational
rules of ordinary quantum theory \cite{Jordan,HaagBannier}. It is of course
true that a nonlinear dynamics of density operators appears highly unnatural
within the linear context of standard quantum kinematics (for instance, on $\Xi(\Hilbert_S)$, 
such dynamics would be represented by bifurcating dynamical trajectories) and
that the causality argument of section \ref{linearityargument1} moreover provides a strong, independent reason for
rejecting such a nonlinear dynamics. But, as already stated in the introduction, 
the possibility is contemplated of a generalized dynamics with standard kinematics 
being a lowest-order approximation to a deeper nonlinear theory, and the aim in this note 
was to investigate the extent to which dynamics is constrained by the kinematics.

\section{Conclusion}
To recapitulate the logical structure of the two arguments for linear quantum 
dynamics; both arguments attempt to derive linearity
from certain conditions, which are easily recognized as in fact being \emph{implied}
by linearity themselves. In the first argument this condition is causality, which may be
viewed as a primitive physical condition, one that must be effectively satisfied in order to 
avoid actual pathologies. It was seen that for causality to imply linearity,
it seems to be necessary to assume that measurements are accompanied by objective 
state reduction, something which is consistent with arguments for linear dynamics
put forward by earlier authors. It was also seen that this actually leads to a 
loophole in the argument, which will presumably be closed by future experiments.
In the view of this author, the strength of this way of arguing for linear dynamics is that 
it relates two concepts which are a priori unrelated, one of which may be regarded as a physical axiom. 
In the second argument, the condition implied by linear dynamics is that 
different proper mixtures belonging to the same initial density operator are
not physically distinguished by the dynamics. It was argued however, that such 
a condition is not a very reasonable one to impose, if generalizations of the standard quantum dynamics are contemplated.

\section*{Acknowledgements}
The author thanks Dennis Dieks for discussions on the foundations of Quantum 
Theory and for providing helpful comments on the draft version of the manuscript.

\bibliographystyle{mltplcit}
\bibliography{lindynv1}

\begin{mcbibliography}{10}
\newcommand{\enquote}[1]{``#1''}

\bibitem{Simon1}
Simon, C., Bu\v{z}ek, V. and Gisin, N. (2001), \enquote{The No-Signaling
  Condition and Quantum Dynamics,} Phys. Rev. Lett., {\bf 87}(170405):1--4, see
  also ArXiv : \texttt{quant-ph/0102125}\relax
\relax
\bibitem{Wald}
Wald, R.~M. (1980), \enquote{Quantum Gravity and Time Irreversibility,} Phys.
  Rev., {\bf D21}(10):2742--2755\relax
\relax
\bibitem{Gisin1}
Gisin, N. (1989), \enquote{Stochastic Quantum Dynamics and Relativity,} Helv.
  Phys. Act., {\bf 62}:363--371\relax
\relax
\bibitem{Bona}
Bona, P. (2003), \enquote{Comment on ``No-Signaling Condition and Quantum
  Dynamics'',} Phys. Rev. Lett., {\bf 90}(208901)\relax
\relax
\bibitem{Jordan2}
Jordan, T.~F. (2006), \enquote{Assumptions That Imply Quantum Dynamics Is
  Linear,} Phys. Rev., {\bf A73}(022101):1--4, see also ArXiv :
  \texttt{quant-ph/0508092}\relax
\relax
\bibitem{Hughston}
Hughston, L.~P., Jozsa, R. and Wootters, W.~K. (1993), \enquote{A Complete
  Classification of Quantum Ensembles Having a Given Density Matrix,} Phys.
  Lett., {\bf A183}:14--18\relax
\relax
\bibitem{dEspagnat1}
d'Espagnat, B., \enquote{An Elementary Note about Mixtures,} in
  \enquote{Preludes in Theoretical Physics,} pp. 185--191 (Amsterdam:
  North-Holland, 1966)\relax
\relax
\bibitem{dEspagnat2}
d'Espagnat, B., {\em Conceptual Foundations of Quantum Mechanics\/} (Reading,
  MA: Addison-Wesley, 1989), second edn.\relax
\relax
\bibitem{Peres}
Peres, A. (1990), \enquote{Neumark's Theorem and Quantum Inseparability,}
  Found. Phys., {\bf 20}:1441--1453\relax
\relax
\bibitem{Bell1}
Bell, J.~S., \enquote{Quantum Mechanics for Cosmologists,} in
  \enquote{Speakable and Unspeakable in Quantum Mechanics,} pp. 117--138
  (Cambridge: Cambridge University Press, 1987)\relax
\relax
\bibitem{Bell2}
Bell, J.~S. (1990), \enquote{Against Measurement,} Physics World, {\bf
  8}:33--40\relax
\relax
\bibitem{ConwayKochen}
Conway, J. and Kochen, S. (2006), \enquote{The Free Will Theorem,} Found.
  Phys., {\bf 36}:1441--1473, see also ArXiv : \texttt{quant-ph/0604079}\relax
\relax
\bibitem{Bub}
Bub, J., {\em Interpreting the Quantum World\/} (Cambridge: Cambridge
  University Press, 1997)\relax
\relax
\bibitem{Zurek3}
Zurek, W.~H. (1991), \enquote{Decoherence and the Transition from Quantum to
  Classical,} Physics Today, {\bf 44}(10):36--44\relax
\relax
\bibitem{Zurek1}
Zurek, W.~H. (1981), \enquote{Pointer Basis of Quantum Apparatus : Into What
  Mixture Does the Wave Packet Collapse ?} Phys. Rev., {\bf
  D24}:1516--1525\relax
\relax
\bibitem{Zurek2}
Zurek, W.~H. (1982), \enquote{Environment-Induced Superselection Rules,} Phys.
  Rev., {\bf D26}:1862--1880\relax
\relax
\bibitem{JoosZeh}
Joos, E. and Zeh, H.~D. (1985), \enquote{The Emergence of Classical Properties
  Through Interaction with the Environment,} Z. Phys., {\bf B59}:223--243\relax
\relax
\bibitem{Adler}
Adler, S.~L. (2003), \enquote{Why Decoherence Has Not Solved the Measurement
  Problem: A Response To P.W. Anderson,} Stud. Hist. Phil. Mod. Phys., {\bf
  4}:135--142, see also ArXiv : \texttt{quant-ph/0112095}\relax
\relax
\bibitem{BassiGhirardi1}
Bassi, A. and Ghirardi, G. (2000), \enquote{A General Argument Against the
  Universal Validity of the Superposition Principle,} Phys. Lett., {\bf
  A275}:373--381, see also ArXiv : \texttt{quant-ph/0009020}\relax
\relax
\bibitem{Kent4}
Kent, A. (1990), \enquote{Against Many-Worlds Interpretations,} Int. J. Mod.
  Phys., {\bf A5}:1745--1762, see also ArXiv : \texttt{gr-qc/9703089}\relax
\relax
\bibitem{Stein}
Stein, H. (1984), \enquote{The Everett Interpretation of Quantum Mechanics :
  Many Worlds or None ?} No\^{u}s, {\bf 18}:635--652\relax
\relax
\bibitem{Squires1}
Squires, E.~J. (1990), \enquote{On an Alleged ``Proof'' of the Quantum
  Probability Law,} Phys. Lett., {\bf A145}:67--68\relax
\relax
\bibitem{Barnumetal}
Barnum, H., Caves, C.~M., Finkelstein, J., Fuchs, C.~A. and Schack, R. (2000),
  \enquote{Quantum Probability from Decision Theory ?} Proc. Roy. Soc. Lond.,
  {\bf A456}:1175--1182, see also ArXiv : \texttt{quant-ph/9907024}\relax
\relax
\bibitem{SchlosshauerFine}
Schlosshauer, M. and Fine, A. (2005), \enquote{On Zurek's Derivation of the
  Born Rule,} Found. Phys., {\bf 35}:197--213, see also ArXiv :
  \texttt{quant-ph/0312058}\relax
\relax
\bibitem{Simon2}
Simon, C., Bu\v{z}ek, V. and Gisin, N. (2003), \enquote{Reply to Bona,} Phys.
  Rev. Lett., {\bf 90}(208902)\relax
\relax
\bibitem{Kent1}
Kent, A. (2005), \enquote{Nonlinearity Without Superluminality,} Phys. Rev.,
  {\bf A72}:012108, see also ArXiv : \texttt{quant-ph/0204106}\relax
\relax
\bibitem{Jordan}
Jordan, T.~F. (1993), \enquote{Reconstructing a Nonlinear Dynamical Framework
  for Testing Quantum Mechanics,} Ann. Phys., {\bf 225}(83-113)\relax
\relax
\bibitem{Wald2}
Wald, R.~M., {\em General Relativity\/} (Chicago: University of Chicago Press,
  1984)\relax
\relax
\bibitem{Kent3}
Kent, A. (2005), \enquote{Causal Quantum Theory and the Collapse Locality
  Loophole,} Phys. Rev., {\bf A72}:012107, see also ArXiv :
  \texttt{quant-ph/0204104}\relax
\relax
\bibitem{Aspectetal}
Aspect, A., Dalibard, J. and Roger, G. (1982), \enquote{Experimental Tests of
  Bell's Inequalities Using Time-Varying Analyzers,} Phys. Rev. Lett., {\bf
  49}:1804--1807\relax
\relax
\bibitem{Weihsetal}
Weihs, S., Jennewein, T., Simon, C., Weinfurter, H. and Zeilinger, N. (1998),
  \enquote{Violation of Bell's Inequality Under Strict Einstein Locality
  Conditions,} Phys. Rev. Lett., {\bf 81}:5039--5043, see also ArXiv :
  \texttt{quant-ph/9810080}\relax
\relax
\bibitem{Preskill}
Preskill, J. (1998), \enquote{Lecture Notes on Quantum Computation,} {\ttfamily
  http://www.theory.cal- tech.edu/people/preskill/ph229/\#lecture}\relax
\relax
\bibitem{BreuerPetruccione}
Breuer, H.-P. and Petruccione, F., {\em The Theory of Open Quantum Systems\/}
  (Oxford: Oxford University Press, 2002)\relax
\relax
\bibitem{Davies}
Davies, E.~B., {\em Quantum Theory of Open Systems\/} (London: Academic Press,
  1976)\relax
\relax
\bibitem{Kent5}
Kent, A., \enquote{Quantum Nonlinearity Revisited,} Perimeter Institute
  Seminar, August 1 2002\relax
\relax
\bibitem{Ferrero}
Ferrero, M., Salgado, D. and S{\'a}nchez-G{\'o}mez, J.~L. (2004),
  \enquote{Nonlinear Quantum Evolution Does Not Imply Supraluminal
  Communication,} Phys. Rev., {\bf A70}(014101):1--4\relax
\relax
\bibitem{JoShaSu}
Jordan, T.~F., Shaji, A. and Sudarshan, E. C.~G. (2004), \enquote{Dynamics of
  Initially Entangled Open Quantum Systems,} Phys. Rev., {\bf A70}:052110, see
  also ArXiv : \texttt{quant-ph/0407083}\relax
\relax
\bibitem{Lindblad}
Lindblad, G. (1976), \enquote{On the Generators of Quantum Dynamical
  Semigroups,} Commun. Math. Phys., {\bf 48}:119--130\relax
\relax
\bibitem{Pearle1}
Pearle, P. (1976), \enquote{Reduction of the State Vector by a Nonlinear
  Schr\"odinger Equation,} Phys. Rev., {\bf D13}:857--868\relax
\relax
\bibitem{Pearle2}
Pearle, P. (1989), \enquote{Combining Stochastic Dynamical State-Vector
  Reduction With Spontaneous Localization,} Phys. Rev., {\bf A
  39}:2277--2289\relax
\relax
\bibitem{GRW}
Ghirardi, G.~C., Rimini, A. and Weber, T. (1986), \enquote{Unified Dynamics for
  Microscopic and Macroscopic Systems,} Phys. Rev., {\bf D34}:470--491\relax
\relax
\bibitem{GPR}
Ghirardi, G.~C., Pearle, P. and Rimini, A. (1990), \enquote{Markov Processes in
  Hilbert Space and Continuous Spontaneous Localization of Systems of Identical
  Particles,} Phys. Rev., {\bf A42}:78--89\relax
\relax
\bibitem{Diosi}
Diosi, L. (1989), \enquote{Models for Universal Reduction of Macroscopic
  Quantum Fluctuations,} Phys. Rev., {\bf A40}:1165--1174\relax
\relax
\bibitem{Percival}
Percival, I.~C. (1994), \enquote{Primary State Diffusion,} Proc. Roy. Soc.
  Lond., {\bf A447}:189--209\relax
\relax
\bibitem{BassiGhirardi}
Bassi, A. and Ghirardi, G.~C. (2003), \enquote{Dynamical Reduction Models,}
  Phys. Repts., {\bf 379}:257--426, see also ArXiv :
  \texttt{quant-ph/0302164}\relax
\relax
\bibitem{Hawking}
Hawking, S.~W. (1976), \enquote{Breakdown of Predictability in Gravitational
  Collapse,} Phys. Rev., {\bf D14}(10):2460--2473\relax
\relax
\bibitem{Hartle1}
Hartle, J.~B. (1994), \enquote{Unitarity and Causality in Generalized Quantum
  Mechanics for Nonchronal Spacetimes,} Phys. Rev., {\bf D49}(12):6543--6555,
  see also ArXiv : \texttt{gr-qc/9309012}\relax
\relax
\bibitem{Hartle2}
Hartle, J.~B., \enquote{Generalized Quantum Theory in Evaporating Black Hole
  Spacetimes,} in R.~Wald (Ed.), \enquote{Black Holes and Relativistic Stars,}
  pp. 195--219 (Chicago: University of Chicago Press, 1998), see also ArXiv :
  \texttt{gr-qc/9705022}\relax
\relax
\bibitem{BaSuPe}
Banks, T., Susskind, L. and Peskin, M.~E. (1984), \enquote{Difficulties for the
  Evolution of Pure States into Mixed States,} Nucl. Phys., {\bf
  B244}:125--134\relax
\relax
\bibitem{UnruhWald}
Unruh, W.~G. and Wald, R.~M. (1995), \enquote{On Evolution Laws Taking Pure
  States to Mixed States in Quantum Field Theory,} Phys. Rev., {\bf
  D52}:2176--2182, see also ArXiv : \texttt{hep-th/9503024}\relax
\relax
\bibitem{Weinberg1}
Weinberg, S. (1989), \enquote{Testing Quantum Mechanics,} Ann. Phys., {\bf
  194}:336--386\relax
\relax
\bibitem{Weinberg2}
Weinberg, S., \enquote{Particle States as Realizations (Linear or Nonlinear) of
  Spacetime Symmetries,} in Y.~S. Kim and W.~W. Zachary (Eds.),
  \enquote{Proceedings of the International Symposium on Spacetime Symmetries,}
  vol.~6 of {\em Nucl. Phys. B (Proc. Suppl.)\/}, pp. 67--75 (1989)\relax
\relax
\bibitem{Weinberg3}
Weinberg, S. (1989), \enquote{Precision Tests of Quantum Mechanics,} Phys. Rev.
  Lett., {\bf 62}(5):485--488\relax
\relax
\bibitem{BrodyHughston1}
Brody, D.~C. and Hughston, L.~P. (2000), \enquote{Information Content for
  Quantum States,} J. Math. Phys., {\bf 41}:2586--2592, see also ArXiv :
  \texttt{quant-ph/9906085}\relax
\relax
\bibitem{BrodyHughston2}
Brody, D.~C. and Hughston, L.~P. (2001), \enquote{Geometric Quantum Mechanics,}
  J. Geom. Phys., {\bf 38}:19--53, see also ArXiv :
  \texttt{quant-ph/9906086}\relax
\relax
\bibitem{Gisin2}
Gisin, N. (1990), \enquote{Weinberg's Non-Linear Quantum Mechanics and
  Superluminal Communications,} Phys. Lett., {\bf A143}(1):1--2\relax
\relax
\bibitem{Polchinski}
Polchinski, J. (1991), \enquote{Weinberg's Nonlinear Quantum Mechanics and the
  Einstein-Podolsky-Rosen Paradox,} Phys. Rev. Lett., {\bf
  66}(4):397--400\relax
\relax
\bibitem{HaagBannier}
Haag, R. and Bannier, U. (1978), \enquote{Comments on Mielnik's Generalized
  (Non Linear) Quantum Mechanics,} Commun. Math. Phys., {\bf 60}:1--6\relax
\relax
\end{mcbibliography}

\end{document}